\documentclass{Interspeech2024}

\interspeechcameraready

\title{Improving Streaming Speech Recognition With Time-Shifted Contextual Attention And Dynamic Right Context Masking}
\name[affiliation={1}]{Khanh}{Le}
\name[affiliation={2,3}]{Duc}{Chau}
\address{
  $^1$Zalo AI, Vietnam\\
  $^2$University of Science, Vietnam\\
  $^3$Vietnam National University, Ho Chi Minh City
}
\email{khanhld218@gmail.com, ctduc@fit.hcmus.edu.vn}


\begin{document}

\maketitle

\begin{abstract}
Chunk-based inference stands out as a popular approach in developing real-time streaming speech recognition, valued for its simplicity and efficiency. However, because it restricts the model's focus to only the history and current chunk context, it may result in performance degradation in scenarios that demand consideration of future context. Addressing this, we propose a novel approach featuring Time-Shifted Contextual Attention (TSCA) and Dynamic Right Context (DRC) masking. Our method shows a relative word error rate reduction of 10 to 13.9\% on the Librispeech dataset with the inclusion of in-context future information provided by TSCA. Moreover, we present a streaming automatic speech recognition pipeline that facilitates the integration of TSCA with minimal user-perceived latency, while also enabling batch processing capability, making it practical for various applications.
\end{abstract}

\section{Introduction}
Connectionist Temporal Classification (CTC) \cite{CTCLUSDWRNN,baevski2020wav2vec}, Attention-based Encoder-Decoder (AED) \cite{E2EABLVSR, chorowski2015attention, 7472621}, and  RNN-Transducer (RNN-T) \cite{STWRNN, zhang2020transformer} are three key approaches in End-to-End Automatic Speech Recognition (E2E ASR). While AED boasts impressive performance with cross-attention between language and acoustic features, it lacks inherent streaming capability, requiring an additional alignment step as in systems like Mocha \cite{MoCha}. In contrast, CTC and RNN-T excel in both non-streaming and streaming applications. 

Contemporary advancements in E2E ASR rely on foundational structures like the Transformer \cite{AIAYN}, Conformer \cite{gulati2020conformer}, and their derivative models \cite{peng2022branchformer, kim2022squeezeformer}. Streaming applications transition from standard convolution and full attention to causal convolution and chunk attention \cite{Chen2020DevelopingRS}. However, this shift leads to performance decline as it struggles to capture a broader acoustic context, a trade-off well-known for balancing latency and accuracy. Therefore, several aspects of streaming capability, including historical context, attention masking techniques, training methods (self/semi-supervised, knowledge distillation), and model architectures, have been extensively studied.
\begin{figure}[ht]
  \centering
  \includegraphics[width=0.88\linewidth]{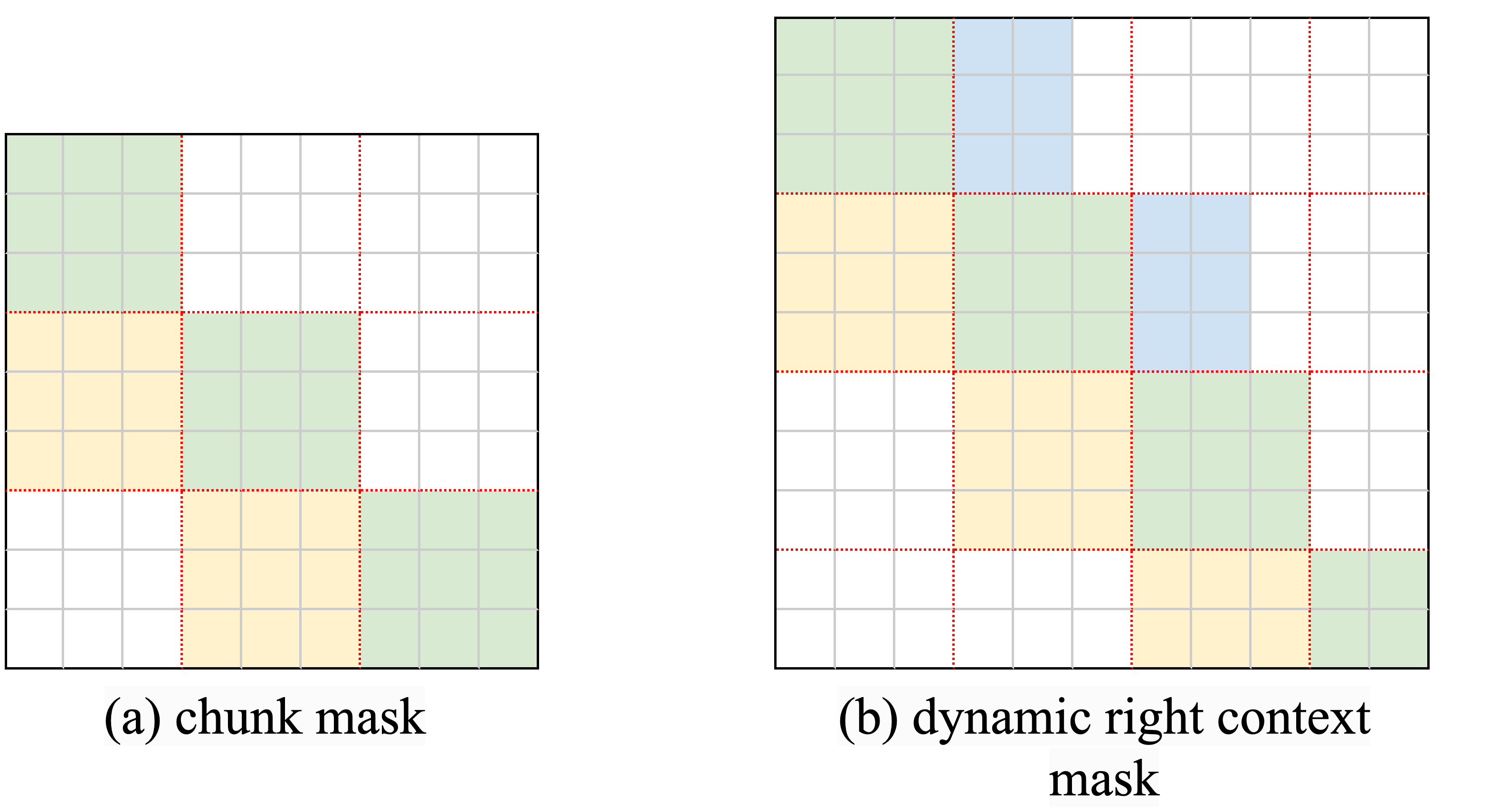}
\vspace{-5pt}
  \caption{(a) represents the conventional chunk mask and (b) is our proposal dynamic right context mask. The areas in yellow, green, and blue are the left context $l$, chunk $c$, and right context $r$, respectively. Here, $l = 3$, $c = 3$ and $r = 2$, in frame units.}

  \vspace{-15pt}
\end{figure}
\setlength{\belowcaptionskip}{5pt} 
For example, \cite{Li2022ImprovingSE} focuses on historical information for streaming models, proposing revisions and updates to the previous output state while the model generates output causally. Their approach, however, increases computational costs due to the need for large revision steps and struggles with batch processing. This becomes especially problematic when revision blocks overlap, causing frames to be decoded multiple times with different contexts each time and thus, requiring an additional alignment step to handle the altered decoding state. 
Besides, in \cite{ISAWNSMDOUD}, knowledge distillation is applied to transfer knowledge from non-streaming to streaming models, employing semi-supervised learning with pseudo-labels generated by a non-streaming model to train the streaming version. Despite its effectiveness, this approach may introduce complexities in training procedures and increase computational demands.
Although the concept of future information has been explored in various studies, latency, primarily stemming from waiting time for real-time future context, still presents a critical challenge in practical applications. Previous work, such as \cite{DBLP:conf/interspeech/AnZOXDW22}, attempted to address this by simulating future context through a GRU simulation encoder. However, their method still falls short of surpassing the performance achieved by utilizing real-time future context, and the simulation encoder requires large-scale datasets to fully demonstrate its effectiveness. Furthermore, \cite{Li2023DynamicCC} introduced Dynamic Chunk Convolution (DCC), successfully applying non-causal convolution to this task. DCC segments audio sequences into chunks, allowing each frame to leverage information from future frames within the current chunk rather than only history frames as in causal convolution. Moreover, the fusion of DCC with Dynamic Chunk Training (DCT) \cite{swietojanski2023variable}, which essentially extends the chunk-based masking to variable chunk sizes, further enhances adaptation to various contexts and boosts the streaming ASR system's performance and flexibility.

In this work, our primary contributions include: 1. Introduction of Time-Shifted Contextual Attention (TSCA) mechanism for real-time creation and revision of in-context future information during decoding, eliminating the need for a simulation encoder as in \cite{DBLP:conf/interspeech/AnZOXDW22}. 2. Development of Dynamic Right Context (DRC) masking technique, which enhances the DCT and DCC methods by dynamically incorporating future information into the attention and convolution layers during training. 3. Designing a streaming pipeline that supports batch processing and efficiently leverages in-context future information from TSCA with minimal user-perceived latency (UPL).

\section{Methods}
\subsection{Time-Shifted Contextual Attention Mechanism}
One drawback of chunk-based attention is that frames at the beginning of the chunk benefit from a larger future context, whereas frames at the end have limited or no access to future context. This limitation can lead to less accurate predictions of the emission token for frames at the chunk's end. Therefore, we hypothesize that providing the trailing frames with future context information may yield greater improvements than merely increasing the chunk size. 
During inference, for each incoming chunk, we cache the previous $r$ input frames and concatenate them with the current chunk frames before feeding them to the model. In the TSCA layer, given the input sequence to attention $x = (\underbrace{\overbrace{x_{-r}, x_{-r+1},..., x_{-1}}^{\text{revised part of size $r$}},\lefteqn{\overbrace{\phantom{x_{0}...,x_{c-r-1},x_{c-r}, x_{c-r+1},...,x_{c-1}}}^{\text{current part of size c}}}x_{0}...,x_{c-r-1}}_{\text{shifted part of size $c$}}, \underbrace{x_{c-r}, x_{c-r+1},...,x_{c-1}}_{\text{in-context future part of size $r$}})$, each output element $z_i$, is computed as a weighted sum of the linearly transformed input element:
\begin{align}
z_i &= \sum_{j=-l_{att}-r}^{c - 1} \alpha^{rel}_{i, j} W_v x_j
\end{align} 
$l_{att}$ is the attention left context size and the negative index refers to frames associated with the preceding chunk. Instead of waiting for future frames, shifting left by $r$ turns the last $r$ frames $x_{[c-r:c-1]}$ of the current chunk into the future context of $x_{[-r:c-r-1]}$, which we refer to as in-context future information\footnote{In this paper, ``future context'' and ``right context'' are interchangeable, both referring to the 'in-context future information' from TSCA.}. This shifting technique enables the model to refine the output of the revised part $x_{[-r:-1]}$, which consists of trailing frames of the previous chunk, and results in performance enhancements. The essence of TSCA lies in optimizing context utilization rather than expanding it. For example, instead of opting for a large chunk size of 960 milliseconds (ms), choosing a smaller size of 640 ms along with a left shift of 320 ms to generate in-context future information offers greater advantages, while also maintaining equivalent computational and GPU memory costs. Notably, TSCA is an inference-time technique, there is no need for additional training.

The relative weight coefficient, $\alpha^{rel}_{i, j}$, is computed using a softmax function:
\begin{align}
 \alpha^{rel}_{i,j} &= \frac{\exp e_{i,j}}{\sum_{k = -l_{att} - r}^{c-1} \exp e_{i,k}}
\end{align}
And $e_{i,j}$ is computed using the scaled dot product function:
\begin{align}
 e_{i,j} &= (x^{T}_{i} W^{T}_{q} W_k x_j + x^{T}_{i} W^{T}_{q} W_{\mathcal{R}} \mathcal{R}_{i-j} \nonumber \\ &+ u^T W_k x_j + v^T W_{\mathcal{R}}\mathcal{R}_{i-j})/\sqrt{d_k}
\end{align}
$W_q$, $W_k$, $W_v$, and $W_\mathcal{R}$ denote the query, key, value, and relative positional weight matrices, respectively. The parameters $u \in \mathbb{R}^{d}$ and $v \in \mathbb{R}^{d}$ are trainable as described in \cite{dai-etal-2019-transformer}. $\mathcal{R} \in \mathbb{R}^{L_{\text{max}} \times d}$ represents the relative positional encodings, with the $i^{th}$ row $\mathcal{R}_i$ indicating a relative distance of $i$ between two positions. Here, $d$ signifies the feature dimension. Note that for $j \in [-l_{att}-r, -r-1]$, the computations for $W_k x_j$ and $W_v x_j$ are already completed in the preceding step, while for $j \in [-l_{att}-r + c, -r-1 + c]$, these values will be cached for the next chunk. 

\subsection{Dynamic Right Context Masking} \label{sec:2.2}
\text{}
\vspace{-10pt}
\begin{algorithm}
  \caption{Dynamic Right Context Mask}
    \textbf{Input}: input size $size$, left context size $l$, chunk size $c$, right context size $r$, and probability $p$.
    
    \textbf{Output}: dynamic right context mask
  \begin{algorithmic}[1]
    \State \textbf{assert} $r < l$
    \State mask $\gets$ $zeros$($size$, $size$)
    \State $i \gets 0$
    \While {$i < size$}
        \State $cur\_c \gets c$
        \If{$rand(0, 1) < p$}
          \State $cur\_c \gets c + r$
        \EndIf
        \State $mask[i:i+cur\_c][i - l:i + cur\_c] \gets 1$
        \State $i \gets i + c$
    \EndWhile
    \\Return $mask$
  \end{algorithmic}
\end{algorithm}

We leverage the idea of DCT, enabling the model to process speech input signals with different context sizes during inference. However, with a broad range of chunk sizes and left context sizes, the training step required significantly more time to converge. This issue was especially noticeable with large training datasets, leading to suboptimal convergence. In DRC, we chose a smaller range of options, guided by a selection strategy that fits with our proposed TSCA mechanism. Denote $C = [c_0, c_1,..., c_n]$ and $R = [r_0, r_1,..., r_n]$  are the chunk size and right context ranges, respectively. Then
\begin{align}
    r_i = r_{i-1} + d \\
    c_i = c_0 + r_i
\end{align}
where $c_0$ and $r_0$ are the smallest chunk size and right context size respectively. $n$ is the range size and $d$ is the common difference between consecutive $r_i$ in $R$. We set $c_0 = 10$, $r_0 = 0$, $n = 3$, and $d = 3$ for all the experiments. This selection strategy enhances model robustness by training with diverse chunk sizes, including both overlap and non-overlap scenarios (e.g., $c = 10, r = 3$ versus $c = 13, r = 0$).  The initial value of $0$ for $r_0$ is critical as it guarantees that the model learns non-overlapping cases during training. This occurs in approximately $\frac{1}{n+1}$ of training steps. In our configuration, there is a $25\%$ chance of this happening.

When generating an attention mask for a specific segment, there is a likelihood of $p$ for the segment to extend by additional $r$ frames to the right, thereby overlapping with the next one. Note that for creating a mask, a valid pair $(c, r)$ should satisfy the condition $r < c$. The hyperparameter $p$ plays a crucial role in controlling the expansion of the receptive field in deeper encoder layers. Specifically, there is a probability of $p^m$ that a single chunk can attend to $m$ consecutive following chunks, with $m$ ranging up to the maximum number of encoder layers in the Conformer model. Hence, choosing a value for $p$ that is excessively large may inadvertently result in significant information leakage, while selecting an overly small value could restrict the accessibility of relevant information. In either case, such settings fail to accurately replicate the inclusion of right context during inference within the training phrase. 
Briefly, this masking strategy allows the model to access future context in the attention and DCC layer, emulating lookahead and thereby mitigating the gap between training and inference when employing TSCA.
\subsection{Dynamic Chunk Convolution with Lookahead} \label{sec:2.3}
We also incorporate right context in DCC, offering flexibility in choosing the decoding chunk size as either $c$ or $c + r_{\text{min}}$, where $r_{\text{min}}$ is equal to $\min(l_{\text{conv}}, r)$ and $l_{\text{conv}} = \frac{\text{kernel size} - 1}{2}$. When opting for a decoding chunk size of $c + r_{\text{min}}$, we split the audio into chunks of size $l_{\text{conv}} + c + r_{\text{min}}$ with a stride of $c$. As only certain segments have access to future contexts, we apply a masking mechanism using the proposed mask to exclude the right context of segments that are not selected.

\subsection{Low User-Perceived Latency Streaming Pipeline Design For Future Context Usage}
\vspace{-5pt}
\label{sec:3.3}
\begin{figure}[ht]
  \centering
  \includegraphics[width=0.9\linewidth]{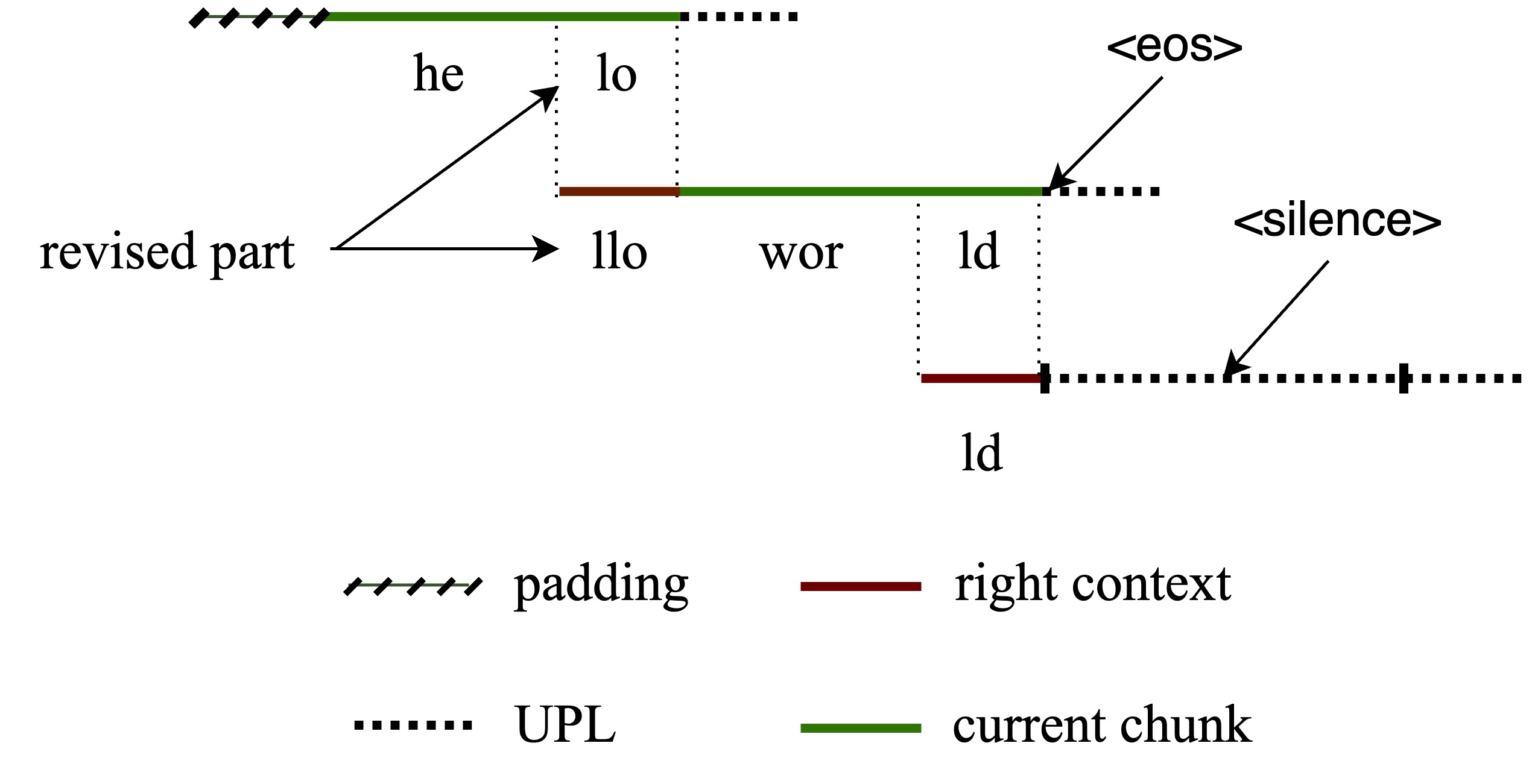}
    \vspace{-3pt}
  \caption{Streaming ASR with future context pipeline design.}
  \label{fig:figure3}
  \vspace{-10pt}
\end{figure}
In the domain of streaming ASR, we address the concept of UPL, denoting the time delay between a user's speech input and the system's recognition of it, subsequently delivering a response. The inclusion of future context can influence UPL in three primary ways: 1. The waiting time for right context, 2. The delay in presenting the text response for the right context portion, and 3. The delay occurs towards the end of the speech, lasting for chunk-size time units plus model computation time.

We propose a streaming ASR system pipeline that effectively utilizes future contextual information without introducing delays while enabling batch processing. Initially, to ensure uniform input sizes for batch processing and allow the left shift of TSCA, we pad the initial data chunk with a size of $r$. Although the TSCA completely removes waiting time for future context, computational overhead remains for right context of the last chunk (as seen in Figure \ref{fig:figure3}). Therefore, our pipeline shares similarities with \cite{Li2022ImprovingSE} in creating the illusion of model ``response'' while executing revised operations, however, our focus lies on revising future information instead. Specifically, we will display not only the current chunk's transcription but also the transcription of its associated future part, hence eliminating delay perception and avoiding redundant computations toward the end of the speech. As depicted in Figure \ref{fig:figure3}, the user will sequentially receive text output as follows: \textit{helo} $\rightarrow$ \textit{hello world} instead of \textit{he} $\rightarrow$ \textit{hello wor} $\rightarrow$ \textit{hello world}. We track the decoding process with variable offset $o$, starting at $-r$, and incrementing by $c$ for each chunk, with negative values used to mask out the padding of the first chunk in the TSCA layer.

\section{Experiments} \label{sec:3}
\begin{table}[htbp]
\centering
\caption{\label{tab:table2}Impact of decoding chunk size for DCC on training and inference. Results are evaluated on dev clean with $c=10$.}
\vspace{-3pt}
\resizebox{0.83\columnwidth}{!}{

\begin{tabular}{|c|c|c|c|c|c|}
\hline
\multicolumn{2}{|c|}{\textit{Decoding Chunk Size}} & \multicolumn{3}{c|}{\textbf{$r$}} & \multirow{2}{*}{\centering Average} \\
\cline{1-5} 
\textit{Training} & \textit{Inference} & 3 & 6 & 9  & \\
\hline


\multirow{2}{*}{$c$} 
  & $c$ & \textbf{4.03} & \textbf{3.83} & \textbf{3.76} & \textbf{3.87} \\
\cline{2-6}
  & $c+r$ & 4.09 & 3.84 & 3.81 &  3.91 \\
\hline
\multirow{2}{*}{$c+r$} 
  & $c$ & 4.20 & 3.93 & 3.81 & 3.98 \\
\cline{2-6}
  & $c + r$ & 4.30 & 4.14 & 4.03 & 4.16 \\    
\hline
\end{tabular}
}
\vspace{-10pt}
\end{table}
\subsection{Setups} \label{sec:3.1}
We train two models with the same underlying architecture: one with chunk-based DCT mask \cite{Li2023DynamicCC} serving as the baseline (A), and one with DRC mask (C). (B) and (D) correspond to (A) and (C) respectively but with TSCA employed at inference. \\
\textbf{Model:} We employ CTC-AED as in \cite{yao2021wenet} for our streaming model, consisting of a Shared Encoder, a CTC Decoder, and an Attention Decoder. Within the Shared Encoder, we adopt the Conformer architecture, comprising 12 layers with 256 feature dimensions, 4 self-attention heads, and the feed-forward layer has outputs of 2048 dimensions. A stack of two convolutional sub-sampling layers with a $3 \times 3$ kernel size and a stride of 2 is utilized, resulting in a time shift of 40 milliseconds. The CTC Decoder is composed of a linear layer followed by a softmax layer to predict the token distribution. The Attention Decoder is a shallow bidirectional transformer, consisting of 6 transformer encoder layers with configurations identical to those in the encoder. The total number of parameters in the model is 50M, with the Encoder accounting for approximately 34M parameters. At inference, we exclusively use the CTC Decoder for its streaming capabilities. Training employs a combined CTC and AED loss function, balanced by the hyperparameter $\lambda = 0.3$:
\begin{align}
    L_{total} &= \lambda L_{CTC} + (1 - \lambda) L_{AED}
\end{align} 
\textbf{Dataset:} We conduct experiments on Librispeech, a public English speech corpus, comprising 960 hours of audio in the training set. We tune the model's hyperparameters using the development data (clean and other) and then evaluate the model on the unseen test data (clean and other) to ensure its effectiveness.
\textbf{Training:} Our audio front end utilizes 80-dimensional filter-bank features with 25ms FFT windows and a 10ms frameshift, augmented using Speed-Perturb and SpecAugment techniques. The Byte-Pair Encoding vocabulary consists of 5000 tokens, each associated with a 256-dimensional embedding. All models are trained from scratch using the WeNet 2.0 toolkit \cite{zhang2022wenet} on three NVIDIA RTX A5500 GPUs with mixed precision. Training lasts 150 epochs with the Adam optimizer and a Noam warm-up learning rate scheduler, spanning 15.000 steps with a peak learning rate of 0.001. The final model averages parameters from the last 35 checkpoints.  \\
\textbf{Evaluation:} 
We use word error rate (WER) and relative WER reduction (rWERR) to evaluate our ASR model. We also compute the confidence interval at a significance level of $\alpha = 5\%$, using bootstrapping method with $B = 5000$ iterations for the metric of interest. This involves repeating the following procedure $B$ times to obtain the empirical distribution of metric values: 1. Randomly sample the test set with replacement to get a new set of the same size as the original. 2. Compute the metric on the new set. Finally, for a significance level $\alpha$, compute the $\alpha$ and the $1 - \alpha$ percentiles of the empirical distribution. The result is represented as $\textit{mean}_{[\textit{min}, \textit{max}]}$, where \textit{mean} refers to the metric value on the original test set and the range denotes the lower and upper bounds of the interval. At inference, we decode using prefix beam search with a beam size of 10 and the left context is fixed at 60 sub-sampled frames for all evaluations.
\subsection{Hyperparameter search}
The optimal decoding chunk size in DCC for training and inference with right context is explored in Table \ref{tab:table2}, indicating that using a decoding chunk size of $c$ during both training and inference yields the best overall results compared to other configurations. The results also suggest that regardless of whether DCC is trained with or without right context, inference with right context should use a decoding chunk size of $c$, involving a splitting method to separate the current chunk and future context.

\begin{figure}[t]
  \centering
  \includegraphics[width=0.75\columnwidth]{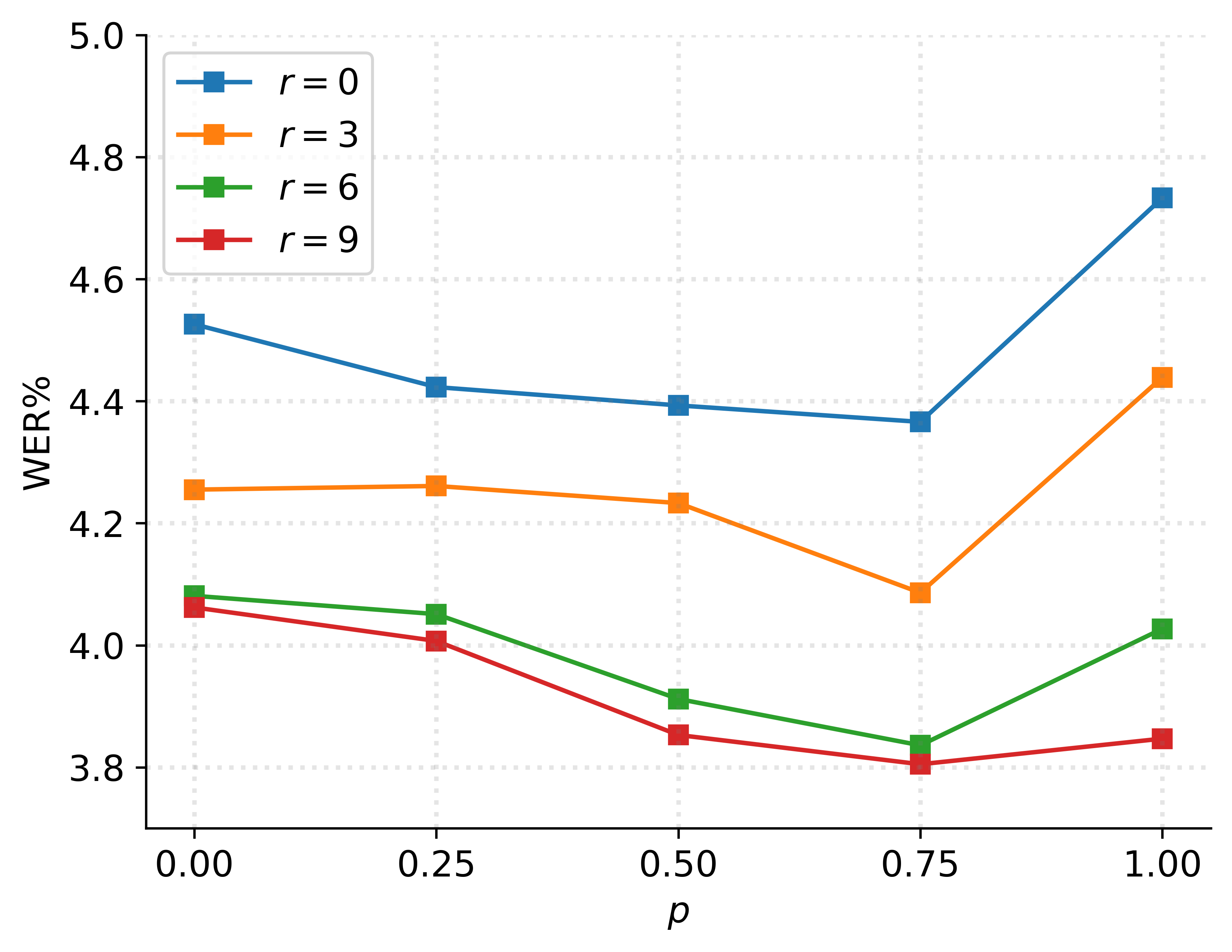}
\vspace{-7pt}
  \caption{WER comparison between different $p$ values with chunk size $c = 10$ on dev clean.}
  \label{fig:figure2}
  \vspace{-22pt}
\end{figure}

We also conducted additional experiments to evaluate the model across various $p$ values. The findings, depicted in Figure \ref{fig:figure2}, reveal a noteworthy trend: as the parameter $p$ increases, the WER consistently decreases, reaching its optimum at $p=0.75$. 
Moreover, as discussed in Section \ref{sec:2.2}, setting $p$ to $1$ resulted in a significant performance drop, especially when little to no future context was available. Similarly, a low value $p = 0.25$ shows minimal improvement or even stagnation in performance.
\vspace{-5pt}
\subsection{Results} \label{sec:3.2}
The findings in Table \ref{tab:table1} highlight a significant enhancement in model performance with the integration of TSCA.
To validate these improvements, we conducted a comparative analysis by computing the rWERR between the average WERs of methods employing TSCA ((B) and (D): $c=[10, 13, 16]$ and $r=[3, 6, 9]$, representing right context sizes ranging from 19\% to 90\% of the chunk size) and those not using TSCA ((A) and (C): $c=[10, 13, 16]$). Subsequently, we utilize the bootstrapping method outlined above to determine the confidence interval of rWERR. The integration of TSCA results in rWERR values of $6.3_{[4.8, 7.7]}\%$ (4.44/4.74) and $5.0_{[4.3, 5.7]}\%$ (11.62/12.23) on test clean and test other when comparing (B) to (A), and $7.4_{[6.2, 8.9]}\%$ (4.24/4.58) and $6.3_{[5.6, 7.2]}\%$ (11.37/12.14) rWERR when comparing (D) to (C), thus, indicating its efficiency. 

Besides, our findings also corroborate our hypothesis, demonstrating that a larger value of \( r \) confers an advantage over a larger \(c\). For instance, when experimenting with \(c=10\) and \(r=[3,6,9]\) (D), the WER is reduced by an average of \(5.6_{[4.3, 7.0]}\%\) (4.22/4.47) on test clean and \(4.0_{[3.3, 4.9]}\%\) (11.39/11.87) on test other over \(c=[13, 16, 19]\) (C). This streaming setup incurs no increase in model computation or GPU memory usage, as the input size remains unchanged. Furthermore, due to our pipeline design, there is no negative impact and even an enhancement on UPL in this scenario. Users could receive text responses more frequently, with text displayed after every 10 frames instead of 13, 16, or 19, provided that the Real Time Factor (RTF) remains below 1. Notably, the RTF when using TSCA is determined by the computational time for processing an input size of \(c+r\) divided by \(c\) time units, not \(c+r\) time units due to the left shift. 

Additionally, our experiments show that the DRC mask, combined with the selection strategy, outperforms the chunk-based mask consistently across all configurations. We evaluate its effectiveness using the same approach as for TSCA. When TSCA is not used ((C) versus (A)), our mask achieves an average rWERR of \(3.4_{[0.6, 5.9]}\%\) (4.53/4.69) on test clean and \(0.9_{[-0.7, 2.7]}\%\) (12.02/12.13) on test other. While in TSCA-enabled scenarios ((D) versus (B)), we observe improvements of \(4.5_{[1.6, 7.3]}\%\) $(4.24/4.44)$ and \(2.2_{[0.3, 4.0]}\%\) $(11.37/11.62)$ on test clean and test other, respectively.

The positive lower boundary values of the confidence interval for rWERR, when employing TSCA and DRC masking, indicate the consistent enhancement of our proposals on streaming capability. Moreover, we observed that all rWERR values in the empirical distribution, obtained from methods employing TSCA against those that either did not utilize TSCA or chose to expand chunk sizes, are positive values, thus emphasizing its robustness across different configurations. Additionally, the DRC mask demonstrates a preference for TSCA, showcasing notable enhancement over DCT with just 0.2\% of negative rWERR values in the empirical distribution and a positive lower bound of the confidence interval. Conversely, in scenarios where TSCA is not utilized, the DRC mask yields a negative lower bound. Nevertheless, only 13.1\% of the elements in the empirical distribution are negative values, indicating that our mask predominantly enhances the baseline model. Finally, we decided to evaluate the streaming ASR system (D) against baseline (A) despite the trade-offs for memory usage and computational time, as both TSCA and chunk-based systems receive equal real-time chunks during inference. As a result, decoding with $c=10$ and $r=6$, results in a relative WERR of $13.9_{[10.8, 16.7]}\%$ (4.16/4.83), and $10.0_{[8.1, 12.0]}\%$ (11.28/12.54) over $c=10$ and $r=0$ on test clean and test other, respectively. While $r=9$ offers potential performance enhancement, we opt for $r=6$, covering 60\% of the chunk size, for a balanced approach.

\begin{table}[t]
\caption{\label{tab:table1}WER results on Librispeech test set.}
\vspace{-5pt}
\centering
\resizebox{\columnwidth}{!}{
    \begin{tabular}{ l c c c c c c c c c   }
    \cmidrule[1pt]{3-10}
    \multicolumn{2}{c}{} & \multicolumn{4}{c}{\textbf{test clean}} & \multicolumn{4}{c}{\textbf{test other}} \\ 
    \cmidrule[1pt]{1-10}
    \multicolumn{1}{c}{\textbf{Method}}  &  \backslashbox{$r \downarrow$}{$c \rightarrow$} &
     10 & 13 & 16 & 19 & 10 & 13 & 16 & 19 \\
    \cmidrule[1pt]{1-10}
    \multirow{3}{*}{\shortstack[l]{(A) Baseline: \\ + DCT}}
        & 0 & 4.83 & 4.69 & 4.70 & 4.53 & 12.54 & 12.20 & 11.94 & 11.85  \\
        \cmidrule(lr){3-5}
        \cmidrule(lr){7-9}
        & & \multicolumn{3}{c}{\textbf{4.74}} & & \multicolumn{3}{c}{\textbf{12.23}} & \\
        \cmidrule(lr){3-6}
        \cmidrule(lr){7-10}
        & & \multicolumn{4}{c}{\textbf{4.69}} & \multicolumn{4}{c}{\textbf{12.13}} \\
    \cmidrule[1pt]{1-10}
    \multirow{4}{*}{\shortstack[l]{(B) Baseline: \\ + DCT \\ + TSCA}}

        & 3 & 4.55 & 4.53 & 4.53 &  & 11.90 & 11.69 & 11.63 & \\
        & 6 & 4.36 & 4.35 &   & & 11.59 & 11.50 &  &  \\
        & 9 & 4.30 &  &   & & 11.38 &  &  & \\

        \cmidrule(lr){3-5}
        \cmidrule(lr){7-9}
        & & \multicolumn{3}{c}{\textbf{4.44}} & & \multicolumn{3}{c}{\textbf{11.62}} & \\
    \cmidrule[1pt]{1-10}
    \multirow{4}{*}{\shortstack[l]{(C) Ours: \\ + DRC}}
        & 0 & 4.71 & 4.63 & 4.41 & 4.38  & 12.47 & 12.15 & 11.81 & 11.64 \\
        \cmidrule(lr){3-5}
        \cmidrule(lr){7-9}
        & & \multicolumn{3}{c}{\textbf{4.58}}  & & \multicolumn{3}{c}{\textbf{12.14}} & \\
        \cmidrule(lr){4-6}
        \cmidrule(lr){8-10}
        & &  & \multicolumn{3}{c}{\textbf{4.47}} &  & \multicolumn{3}{c}{\textbf{11.87}} \\
        \cmidrule(lr){3-6}
        \cmidrule(lr){7-10}
        & & \multicolumn{4}{c}{\textbf{4.53}} & \multicolumn{4}{c}{\textbf{12.02}} \\
    \cmidrule[1pt]{1-10}
    \multirow{5}{*}{\shortstack[l]{(D) Ours: \\ + DRC \\ + TSCA}}

        & 3 & 4.43 & 4.34 & 4.29 &  & 11.88 & 11.58 & 11.38 &  \\
        & 6 & 4.16 & 4.13 &  & & 11.28 & 11.12 & \\
        & 9 & 4.07 &  &  &  & 11.00 &  & \\
        \cmidrule(lr){3-3}
        \cmidrule(lr){7-7}
        & $[10]$ & \textbf{4.22} &  & & & \textbf{11.39} \\
        \cmidrule(lr){3-5}
        \cmidrule(lr){7-9}
        & $[10-16]$ & \multicolumn{3}{c}{\textbf{4.24}} & & \multicolumn{3}{c}{\textbf{11.37}} \\
    \cmidrule[1pt]{1-10}
    \end{tabular}
}
\vspace{-15pt}
\end{table}

\section{Conclusions}
In this paper, we introduced Time-Shifted Contextual Attention and Dynamic Right Context masking, two novel techniques for improving streaming ASR performance by leveraging future context information. Experiments on the Librispeech dataset show significant WER reductions, up to 13.9\% relative to chunk-based model. We also present a low-latency streaming ASR pipeline that integrates TSCA with minimal latency overhead, enabling quick adaptation to varying context sizes. Our findings underscore the importance of considering future context in streaming ASR systems and highlight the potential of our proposed techniques for real-world applications.

\nocite{*}
\bibliographystyle{ieeetran}
\bibliography{mybib}

\end{document}